\documentclass[aps,superscriptaddress,superbib,twocolumn,groupedaddress,showpacs]{revtex4}
\usepackage[final]{neel}

\bibpunct{[}{]}{,}{n}{}{}

\usepackage[latin1]{inputenc}
\usepackage[T1]{fontenc}

\def\figurename{Fig.}

\begin{document}

\renewcommand\topfraction{0.8}
\renewcommand\bottomfraction{0.8}
\renewcommand\floatpagefraction{0.8}

\title{Some aspects of Magnetic Force Microscopy of hard magnetic films}%

\author{G. Ciuta}
\affiliation{Univ. Grenoble Alpes, Inst NEEL, F-38000 Grenoble, France}%
\affiliation{CNRS, Inst NEEL, F-38000 Grenoble, France}

\author{F. Dumas-Bouchiat}
\altaffiliation[Present address: ]{Centre Europ\'{e}en de la C\'{e}ramique, Univ. Limoges, CNRS, France}
\affiliation{Univ. Grenoble Alpes, Inst NEEL, F-38000 Grenoble, France}%
\affiliation{CNRS, Inst NEEL, F-38000 Grenoble, France}

\author{N. M. Dempsey}
\affiliation{Univ. Grenoble Alpes, Inst NEEL, F-38000 Grenoble, France}%
\affiliation{CNRS, Inst NEEL, F-38000 Grenoble, France}

\author{O. Fruchart}
\email[]{Olivier.Fruchart@neel.cnrs.fr}
\affiliation{Univ. Grenoble Alpes, Inst NEEL, F-38000 Grenoble, France}%
\affiliation{CNRS, Inst NEEL, F-38000 Grenoble, France}

\date{\today}

\pacs{07.79.Pk, 75.50.Ww,75.60.Ch}


\begin{abstract}

A number of aspects of magnetic force microscopy (MFM) specific to the imaging of hard magnetic films have been studied. Firstly, we show that topographic images made in tapping mode with probes characterized by the moderate cantilever stiffness usual for MFM~($\unit[1\mathrm{-}4]{\newton\per\metre}$), contain artifacts due to strong probe-sample interactions which lead to probe retraction. As a result, stiffer cantilevers (\eg $\unit[40]{\newton\per\metre}$) are better adapted to characterizing such hard magnetic films. Secondly, imaging with probes coated by a hard magnetic film leads to phase maps which show a two-fold symmetry, with paired dark/light contrast on opposite domain edges along the direction of the cantilever. This is due to the tilt of the direction of tip magnetizaiton and direction of oscillation, with respect to the sample normal. Thirdly, due to the long-range nature of the stray field produced by hard magnetic films containing micron-sized domains, MFM phase contrast reflects the stray field itself, as opposed to that of its spatial derivatives, as is generally the case in MFM.

\end{abstract}

\maketitle



\newcommand{\Fts}{\ifmmode{F_\mathrm{ts}}\else{$F_\mathrm{ts}$}\fi}

\section{Introduction}
\label{sec-introduction}

Rare earth-based permanent magnets are key materials for a variety of applications, being used in devices ranging from small-scale actuators (\eg voice coil motors of hard disk drives) to large scale motors and generators (hybrid electric vehicles, gearless wind turbines). Growing demand for such high performance materials for energy applications, coupled with concerns about the pricing and sourcing of rare earths\cite{bib-GUT2011}, has led to a resurgence in the study of hard magnetic materials. Thin and thick films may serve as model systems for studying the development of coercivity in NdFeB-based magnets, with the choice of preparation conditions offering a certain control over the film's microstructure (grain size, shape and orientation, secondary phases\ldots)\cite{bib-KAP1993,bib-SER2003b,bib-NEU2004,bib-DEM2007}. The very high values of coercivity recently reported in heavy-RE free $\mathrm{Nd}_2\mathrm{Fe}_14\mathrm{B}$-based films\cite{bib-SAT2011,bib-CUI2011,bib-DEM2013} attest of the interest of using films as model systems to study magnetization reversal. Detailed imaging of the magnetic domain structures of such films should contribute to improving our understanding of the complex relationship between microstructure and coercivity.

Magnetic force microscopy~(MFM) was first reported in 1987\cite{bib-MAR1987b,bib-SAE1987}, as a derivation of the then recently developed atomic force microscopy\cite{bib-BIN1982}. MFM offers much potential for the study of the magnetic microstructure of permanent magnets using relatively simple, laboratory based equipment. It's good spatial resolution (routinely $\unit[50]{\nano\metre}$) makes it adapted to the characterization of fine domain structures that cannot be resolved with magneto-optical microscopy.  This is particularly relevant for characterizing the domain structure of the fine grained structures produced in certain bulk materials (e.g. melt-spun ribbons\cite{bib-ALK1998b}, hot-deformed magnets\cite{bib-THI2012b}) and films\cite{bib-NEU2005,bib-CHE2008b,bib-WOO2009}. In the case of film samples, there is no need for time-consuming and potentially damaging sample preparation (such as sample thinning for Lorentz microscopy)\cite{bib-CHA1999}. Specific aspects of imaging permanent magnets have been reported, such as the need for high-coercivity tips to avoid the switching of the tip magnetization in the stray field of the magnet\cite{bib-FOL1996,bib-FOL1998} and image contrast analysis, allowing the distinction between charge and susceptibility contrast\cite{bib-ZUE1998}.

In this manuscript we report on certain aspects of imaging permanent magnet films using MFM. We study the influence of the mechanical stiffness of the cantilever and the magnetic characteristics of the MFM probe, and demonstrate the impact of the probe tilt during oscillation. The origin of image contrast is studied quantitatively on test structures containing magnetic domains of controlled size and shape.

\section{Experimental}
\label{sec-experimental}

Films of out-of-plane textured NdFeB~(thickness $\unit[4]{\micro\metre}$) and isotropic SmCo~(thickness $\unit[2]{\micro\metre}$) with coercivity values of $\unit[2]{\tesla}$ and $\unit[4]{\tesla}$, respectively, were prepared using high-rate triode sputtering\cite{bib-WAL2008,bib-DEM2007}. Ta was used both as buffer and capping layers~($\unit[100]{\nano\metre}$), to avoid sample deterioration during the post-deposition annealing step. Test structures for quantitative analysis were prepared in some films. These consist of magnetic domains of controlled size and shape prepared using the Thermo-Magnetic Patterning~(TMP) technique\cite{bib-DUM2010}. In this technique, the direction of magnetization of a hard magnetic film is switched locally by irradiating (heating) it with a nanosecond pulsed laser through a mask with micron-sized features in the presence of an external magnetic field that is weaker than the film's room temperature value of coercivity. For this study, an out-of-plane magnetised SmCo film was irradiated with a pulsed excimer laser through a mask with an array of square holes of side length $\unit[7]{\micro\metre}$ and pitch $\unit[10]{\micro\metre}$, in a field of $\unit[0.5]{\tesla}$ directed opposite to the film's direction of magnetization. As a result, an array of micro-magnets was obtained, embedded in a matrix with opposite magnetization. Based on a previous quantitative study of the TMP process, we consider that the film is reversed through a depth of circa $\unit[1]{\micro\metre}$\cite{bib-DUM2010}. Magnetic Force Microscopy was performed in air at room temperature with an NT-MDT NTegra-Aura\texttrademark\ instrument. Probes with different mechanical and magnetic characteristics were used. The first set of probes were from Nanosensors\texttrademark, PPP - MFMR, which are typical MFM probes, based on force-modulation cantilevers: $k\simeq\unit[4]{\newton\per\metre}$ and $\omegaZero\simeq\unit[75]{\kilo\hertz}$ coated with $\unit[30\mathrm{-}50]{\nano\metre}$ of $\mathrm{Co}_{80}\mathrm{Cr}_{20}$, tip coercivity $\unit[30]{\milli\tesla}$, hereafter called "flexible cantilever soft coating" probes. Two other types of probes were from Asylum Research\texttrademark. The first type were off-the-shelf force modulation cantilevers~(cantilever ref: AC240TS, probe ref: ASYMFMHC, $k\simeq\unit[3]{\newton\per\metre}$ and $\omegaZero\simeq\unit[60]{\kilo\hertz}$). The second type were customized probes made with stiffer cantilevers, $k\simeq\unit[40]{\newton\per\metre}$ and $\omegaZero\simeq\unit[300]{\kilo\hertz}$, usually used for topography imaging (ref: Olympus AC160TS). The latter two cantilever types were coated with hard magnetic CoPt (nominally, $\muZero\Hc=\unit[0.5]{\tesla}$), and hereafter are called "flexible cantilever hard coating" and "stiff cantilever hard coating" probes, respectively. All probes were magnetized perpendicular to the cantilever prior to imaging, the low coercivity tips in the stray field of a permanent magnet ($\approx\unit[0.5]{\tesla}$), the high coercivity tips in a field of $\unit[7]{\tesla}$ produced by a superconducting coil.

Imaging proceeds with two passes per line in the AC mode with amplitude modulation. The cantilever is excited close to its resonance frequency, and the typical peak-to-peak oscillation magnitude at the tip apex is $\unit[50]{\nano\metre}$. For each line, during the first pass topography is imaged in a light tapping mode to avoid abrasion of the coating, with a setpoint typically less than $\unit[1]{\%}$ below the magnitude measured in the immediate vicinity above the surface. In the second pass the feedback loop is open; the probe is raised by a certain amount $\Delta z$ above the surface and retraces the topography profile of the first pass. We assume that the long-range forces felt during the second pass are purely of magnetic origin. The magnetic image consists of the phase signal during the second pass, with the convention that the signal decreases across resonance: attractive forces result in a negative contrast, depicted as dark contrast in the images. Due to the presence of the Ta capping layer, the actual distance between the apex of the tip and the upper surface of the magnetic material is of the order of $\Delta z=+\unit[100]{\nano\metre}$.

\section{Influence of magnetic coating and cantilever stiffness}
\label{sec-stiffness}

\begin{figure}
  \begin{center}
  \includegraphics[width=82.877mm]{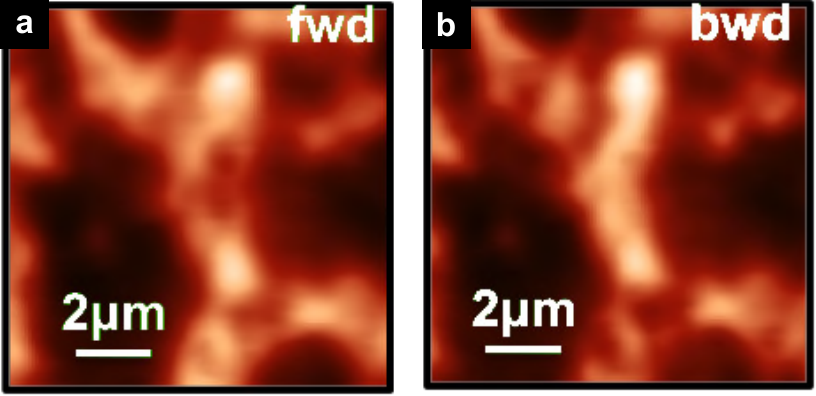}%
  \caption{\label{fig-tip-switching}(a)~Forward and (b)~backward scans of a demagnetized NdFeB film, performed with a "flexible cantilever soft coating" MFM probe.}
  \end{center}
\end{figure}

The significance of the coercivity of the probe material has already been discussed in a report on MFM imaging of permanent magnets\cite{bib-FOL1998}. It was argued that, to ease contrast analysis, the probes should ideally be either very coercive, to remain rigid within the stray field of the sample, or else soft enough to follow its direction at any point, the latter coming at the expense of loosing information on the direction of magnetization. A detailed inspection of MFM images of a partially demagnetized NdFeB film made with a "flexible cantilever soft coating" probe confirms the moderate value of coercivity of these probes: reversal of the magnetization of the tip was observed and detected at different locations and depending on the scan direction\bracketfigref{fig-tip-switching}. So, for the moment let us focus on the use of tips with a hard coating.

\begin{figure}
  \begin{center}
  \includegraphics[width=83.232mm]{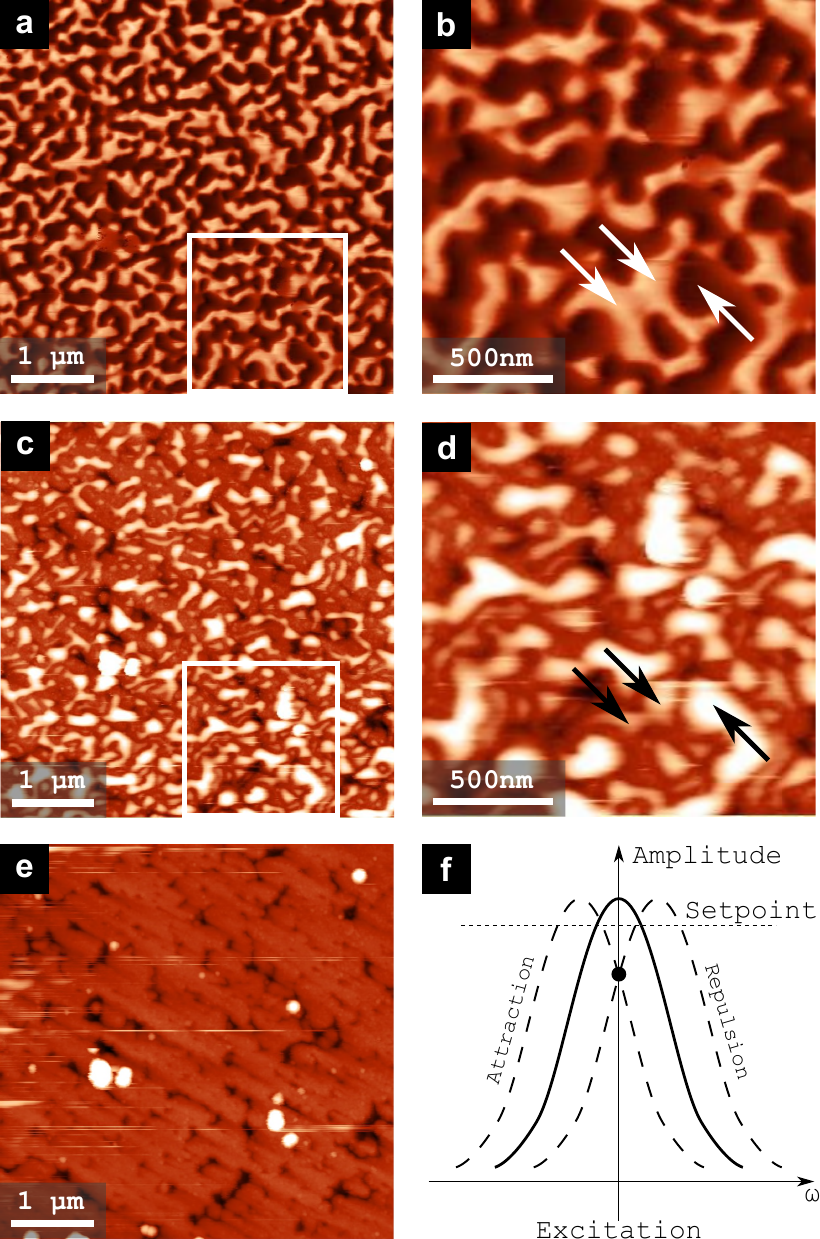}%
  \caption{\label{fig-repulsion-from-surface}(a,b)~MFM images of a NdFeB film with a flexible cantilever hard coating MFM probe, with (c,d) the corresponding topography images, performed under soft tapping conditions. (b,d) are the zoom of the areas marked by a square in (a,c). Downward~(resp. upward) arrows indicate regions with repulsive~(resp. attractive) forces. (e)~Topography of the same area as (a,c), imaged under hard tapping conditions. (f)~Illustration of the shift of resonance peak under attractive and repulsive conditions, lowering the amplitude below the setpoint. }
  \end{center}
\end{figure}

Imaging electric or magnetic forces with an amplitude modulation technique in air requires a compromise between a cantilever stiff enough to image topography reliably and avoid sticking to the surface during the second pass, while flexible enough to allow for a reasonable signal-to-noise ratio in the second pass while monitoring the phase. Cantilevers with stiffness in the range $\unit[1\mathrm{-}4]{\newton\per\metre}$ and resonance frequency $\unit[50\mathrm{-}100]{\kilo\hertz}$ are commonly considered to be a good compromise, and the are the typical commercially available cantilevers proposed for MFM. \figref{fig-repulsion-from-surface} shows MFM~(a,b) and topography~(c,d) images of a NdFeB film in the virgin state, made with a "flexible cantilever hard coating" probe. The magnetic image displays up and down domains of essentially equal total area. Due to the large value of stray field, the force between probe and sample is so large that the phase shift often exceeds $\pm\pi/4$. This is detrimental for several aspects of the MFM imaging process.

Firstly, this compromises the quantitative analysis of the images: even though the phase shift could in principle be converted to a frequency shift to recover linearity with the sensed force, the change of amplitude of oscillation due to the shift of resonance angular frequency changes the probing range and thus the strength of the field as well as the spatial resolution. Note that one may implement a feedback on the amplitude during the second pass, to correct this.

Secondly, when performed in a soft tapping mode to protect the tip apex from mechanical abrasion (setpoint typically about $\unit[1]{\%}$ lower than the free amplitude), topographic images display artifacts due to the probe retracting from the surface by up to several hundreds of nanometers\bracketsubfigref{fig-repulsion-from-surface}{c,d}. Indeed, in the case of strong repulsive or attractive magnetic forces, the shift of resonance angular frequency\bracketsubfigref{fig-repulsion-from-surface}f results in a sizeable decrease of the cantilever amplitude. This decrease is wrongly interpreted by the feedback loop as an increase in tapping interaction with the surface, so the probe is retracted to try to recover a higher amplitude. If the value of free amplitude becomes lower than the setpoint, then the tip retracts from the surface to try to achieve recovery. Note that the static deflection from such forces remains tiny, a few angstr\"{o}ms at most, so that the direct effect of this force can be ruled out as an explanation for retraction from the surface. Besides, in the MFM second pass phase, observation of the areas indicated by arrows on \subfigref{fig-repulsion-from-surface}{b,d} confirms that both repulsive and attractive magnetic forces (due to up and down domains) lead to retraction from the surface. In the present case repulsive forces had a larger impact on topography, because the excitation frequency had been set slightly to the left of the peak maximum. Such artifacts may be avoided by setting an amplitude setpoint well below the free oscillation value, in a so-called hard tapping mode\bracketsubfigref{fig-repulsion-from-surface}e, which however accelerates abrasion of the magnetic material on the probe.

These artifacts could in principle be suppressed through the use of much thinner magnetic coatings, though this may in turn lead to a reduction in the value of coercivity. In the following section we describe experiments made using "stiff cantilever hard coating" probes. In this case, phase shifts are of a few degrees at most (with the same probe coating composition and thickness). As the phase shift scales like $(Q/k)(\nabla F)$, the associated decrease of amplitude scales with $(Q^2/k^2)(\nabla F)^2$~($Q$ is the quality factor of the cantilever, see details in annex~1). Thus, the oscillation amplitude is only reduced by a tiny amount and imaging is faithful. Then concerning abrasion, the extra tip-surface repulsive force in the semi-contact first pass, resulting from the need to use a low setpoint to avoid leaving the surface during second pass, scales as $(Q^2/k)(\nabla F)^2$ for the same amplitude of oscillation as the stiffness of the cantilever is~$k$. With $Q\approx500$ and 200 for the stiff and flexible cantilevers, respectively, while the stiffness varies by a factor of 20, the abrasion force is in the end of the same order of magnitude, though somewhat less for the stiff cantilever~(see Annex~1). Note that decreasing the oscillation amplitude to lower values is also possible without facing mechanical instabilities with the stiff cantilevers. However, in practice, this possibility is limited by the decrease of signal-to-noise ratio for measuring the phase signal. The best compromise may be to use cantilevers with stiffness $\unit[10]{\newton\per\metre}$~(\eg AC200TS, frequency $\unit[115]{kHz}$) and operate them with an oscillating magnitude of the order of $\unit[10\mathrm{-}20]{\nano\metre}$.

\section{Analysis of magnetic contrast}
\label{sec-contrast}

Let us firstly consider the essential physics of contrast analysis. For this purpose, we make simplifying assumptions about the tip and cantilever, which we shall examine in more detail below. As recalled in Annex~1, in magnetic force microscopy, information about the magnetic force \Fts\ acting on the tip and due to the sample, is mapped as the phase shift induced on the cantilever when excited at its natural resonance frequency $\omegaZero$:

\begin{equation}%
\label{eqn-phaseShift}
  \Delta\Phi=-\frac{Q}{k}\fracdiff{\Fts}{z}
\end{equation}
where $z$ is the coordinate of the tip apex along the direction of oscillation. The difficulty in the analysis of magnetic contrast lies in the modeling of \Fts. We restrict the discussion to the case of perfectly rigid sample and tip magnetization. This assumption is valid for hard-magnetic samples and tips as studied here. We provide in Annex~2 the mathematical modeling of this case, while we discuss here its physical grounds.

\begin{figure}
  \begin{center}
  \includegraphics[width=85.885mm]{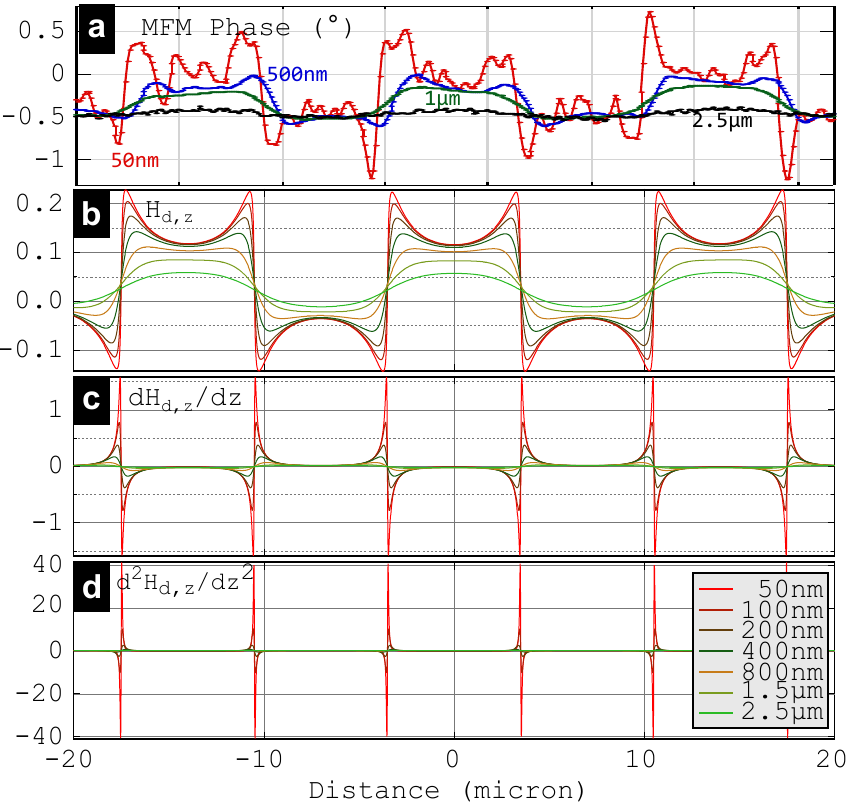}%
  \caption{\label{fig-variousHeights}(a)~Experimental line scans of the phase shift, averaged over a width of 50 lines, and measured at different scan heights above the TMP SmCo film. Note that $\unit[100]{\nano\meter}$ should be added to these values to get the true lift height above the magnetic material, due to the thickness of the Ta capping layer (b-d)~calculated line profiles of $H_{\mathrm{d},z}$, $\linediff{H_{\mathrm{d},z}}{z}$  and $\linediffn{H_{\mathrm{d},z}}{z}2$, expressed in unit of $\Ms$ (A/m), $\Ms/\micro\metre$ ($\unit[10^6]{\ampere\per\metre\squared}$) and $\Ms/\micro\metre^2$ ($\unit[10^{12}]{\ampere\per\metre\cubed}$),  respectively. The lift height is the true lift height.}
  \end{center}
\end{figure}

Magnetically-rigid tips are modeled in the literature as a magnetic dipole or monopole, the phase shift then reflecting the second or first derivative in $z$ of the sample stray field, respectively\cite{bib-HAR1989}. Based on MFM measurements made at various heights in the well-controlled magnetic field produced by current rings, it was concluded that both models could fit experimental data. The magnetic charge for a monopole, and the magnetic moment for a dipole, as well as their fictive position along the tip's axis, can be determined through calibration procedures\cite{bib-LOH1999}. We followed the latter procedure by mapping the phase shift at various lift heights above the TMP-patterned SmCo film consisting of an array of $\unit[7]{\micro\metre}$-edge square magnetic dots in an oppositely-magnetized matrix~\bracketfigref{fig-variousHeights}. For these experiments we used the "stiff cantilever hard coating" probes. We assume at first that the oscillation direction $z$ is normal to the sample plane, and that the magnetic material capping the tip is magnetized along~$z$. We will examine these two conditions later on. The experimental curves are compared with the vertical component of the calculated stray field, and with its first two space derivatives along this direction\bracketfigref{fig-variousHeights}. These were calculated for a depth of reversal of $\unit[1]{\micro\metre}$ in the TMP sample , however changes in this value have very little impact on the curves (the topmost surface charges are dominating the contrast). Thus, the comparison between experiments and theory is robust against uncertainties on the exact depth of reversed magnetization in the TMP process. Note also that to compare experimental results with simulations, one should add the thickness of the Ta capping layer~(\unit[100]{\nano\metre}) to the lift height set by the microscope.

The contrast may be viewed as consisting of two contributions: one part is essentially uniform above the dots and the matrix, their respective intensities being of opposite sign, while the other part is a sharp contrast at the perimeter of the dots.  The latter is related to the expected singularity of magnetic stray fields at sharp edges of prisms. Inspection of the ratio of the magnitude of the two contributions is instructive. It is clear that the experimental contrast is best reproduced by the map of the stray field itself, not by either of its derivatives, the second derivative being the least convincing. This is in sharp contrast with literature reports on MFM, which we discuss below.

It can be demonstrated that mapping the stray field is expected for a tip modeled by a cone uniformly capped with magnetic material~(see Annex~2). In practice, this may be obtained by a directional deposition flux along the tip axis. While this is the case for most commercial probes, in practice the short range of stray fields produced by most samples leads to a cut-off in the magnetic probing volume, resulting in a more localized distribution of the tip magnetic matter contributing to the force. Due to the cut-off, the magnetic coating is better described with a monopole or dipole, or a combination of the two. This effect is amplified by the finite radius of curvature of the tip apex combined with the directional nature of the deposition techniques usually used to deposit the magnetic material, resulting in a larger thickness of material at the apex than on the sides of the tip cone. The direct probing of the stray field is therefore an aspect specific to permanent magnets displaying domains with large lateral size and depth, which give rise to long-range stray fields.

\begin{figure}
  \begin{center}
  \includegraphics[width=85.899mm]{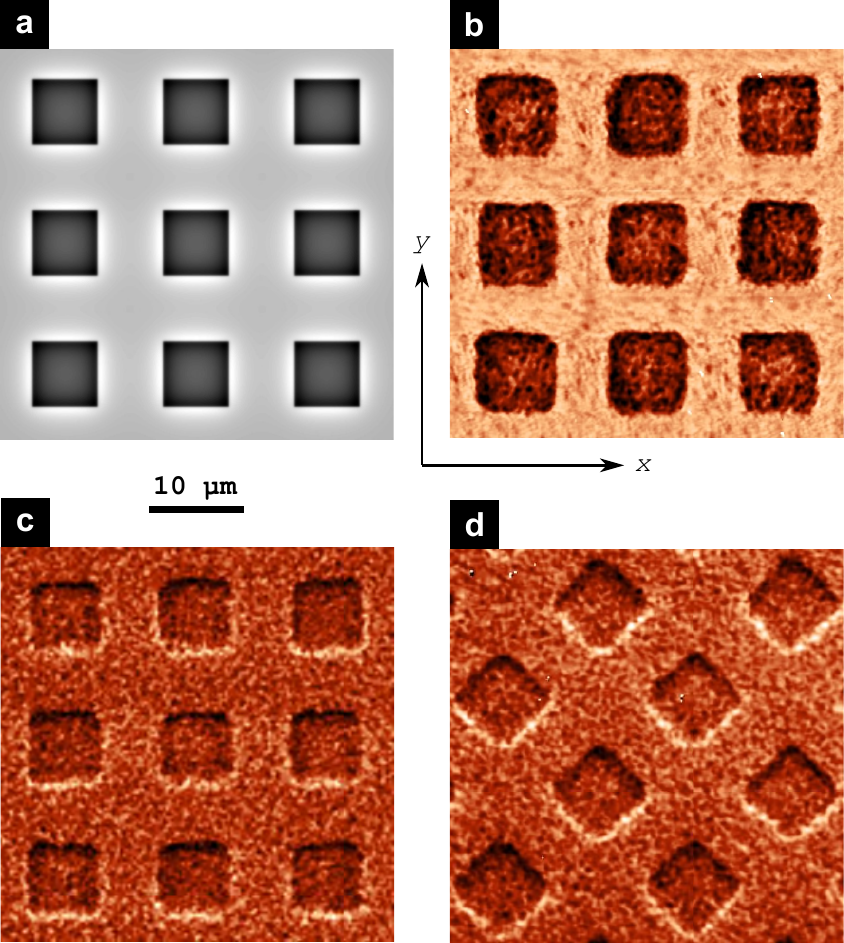}%
  \caption{\label{fig-asymmetricContrast}(a)~Calculated map of the vertical component of the stray field at a height of $\unit[100]{\nano\metre}$ above the patterned SmCo film; (b) phase shift map measured with a "flexible cantilever soft coating" probe; (c,d) phase shift maps measured with a "stiff cantilever hard coating" probe. The sample was rotated through an angle of roughly $\angledeg{45}$ between (c) and (d). }
  \end{center}
\end{figure}

The analysis of the phase maps may be refined further. Irrespective of the tip model discussed above, phase maps should display a four-fold symmetry with dark/light contrast on all four sides of each prism, as calculations show in \subfigref{fig-asymmetricContrast}a.

\begin{figure}
  \begin{center}
  \includegraphics[width=60mm]{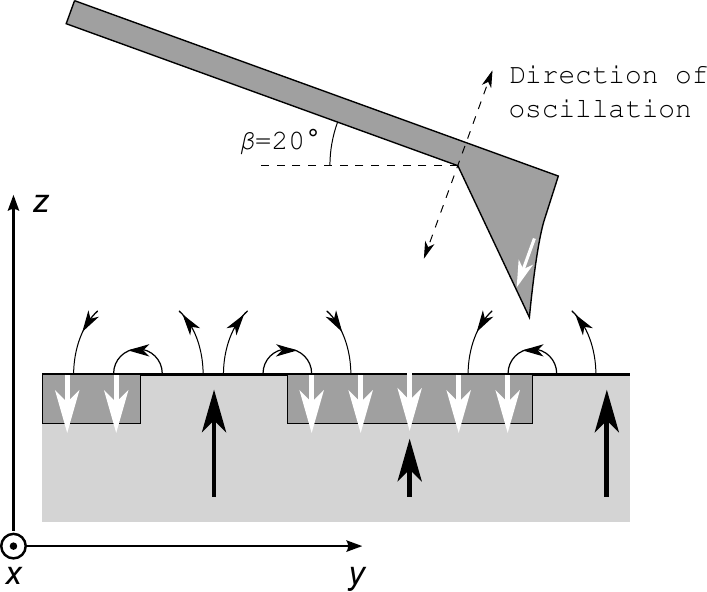}%
  \caption{\label{fig-tiltedTip}Side view schematic of the tilted cantilever.}
  \end{center}
\end{figure}

Experimental phase maps made with a "stiff cantilever hard coating" MFM probe show a clear two-fold symmetry, with paired dark/light contrast on opposite edges along the $y$ direction, and a much weaker contrast along the $x$ direction\bracketsubfigref{fig-asymmetricContrast}c. Dark~/ bright edges were also evidenced in MFM images of the domain structure  in hot-deformed magnets, though they were not explicitly discussed\cite{bib-THI2012}. Before discussing the possible reasons for this, it should be noted that the experimental scans present a fine structure that is not found in calculations. This is a consequence of the granular and isotropic nature of the film. Local stray fields produced by individual grains are detected close to the film surface, but not far from it (see below). Such fine contrast is also found in textured NdFeB films, though it is weaker, owing to the much lower misalignment of grains with respect to the film normal. Returning to the discussion on symmetry, one could argue that the asymmetry of contrast may result from the micro-magnets displaying an in-plane component of magnetization along $y$, resulting for instance from a misalignment of the external magnetic field during the TMP process. This is however ruled out by imaging the same sample rotated by an arbitrary in-plane angle, and still detecting strong edge contrast along the $y$ direction\bracketsubfigref{fig-asymmetricContrast}d. It should also be noted that the contrast does not depend on the choice of the fast and slow scan directions, along $x$ or~$y$. The reason must thus come from the MFM probe or measurement itself. The role of the probe in producing such asymmetric features was first discussed in detail by Rugar \etal, resulting from both the tilted direction of tip magnetization, and tilted direction of oscillation\cite{bib-RUG1990}. The first effect is related to a component of the probe magnetization being off-axis, thus with a component in a given direction in-the-plane of the sample. In this case, the probe would be sensitive to the component of the sample's stray field along this in-plane direction. A random orientation of this component can be ruled out in our case as the asymmetry was always observed along the $y$ direction, whereas fluctuations from probe to probe would be expected for polycrystalline coatings. A systematic effect may come from the fact that, as in most microscopes and because of space constraints, the cantilever is tilted with respect to the sample plane by an angle~$\beta$~($\approx\angledeg{20}$ in our setup, see \figref{fig-tiltedTip}). As the tips are magnetized with a field perpendicular to the cantilever, the moment of the tip is expected to along an angle $\beta$ with respect to the sample normal. A complication is that for the AC240 and AC160 probe series the tip axis is not perpendicular to the cantilever, with a potential impact on the anisotropy of grains at the tip apex. The second effect is related to the tilted direction of oscillation, for the same geometrical reason as mentioned above.

In the monopole model for example~(see Annex~1 and \eg \citenum{bib-RUG1990}), tilting the probe results in sensitivity to i)~gradients of both out-of-plane and in-plane components of the stray field, since part of the tip moment is in the plane of the sample (even though it may be aligned along the tip cone axis, the latter being tilted) and ii)~the planar gradient of any of the field components probed. As regards the first aspect, cantilevers are tilted along~$y~$ with the angle $\beta=\angledeg{20}$ in NT-MDT microscopes\bracketfigref{fig-tiltedTip}. For hard magnetic coatings, the direction of magnetization of the probe may be assumed to be parallel to the magnetizing field, which we apply perpendicular to the cantilever. Thus, the moment of the probe makes an angle $\beta$ forward~$-y~$. As regards the second aspect, since we have shown that MFM images of permanent magnets reflect primarily the stray field itself, gradients do not appear in the phase. As a consequence, we should map $H_{\mathrm{d},z}\cos\beta - H_{\mathrm{d},y}\sin\beta$.

\begin{figure}
  \begin{center}
  \includegraphics[width=85.699mm]{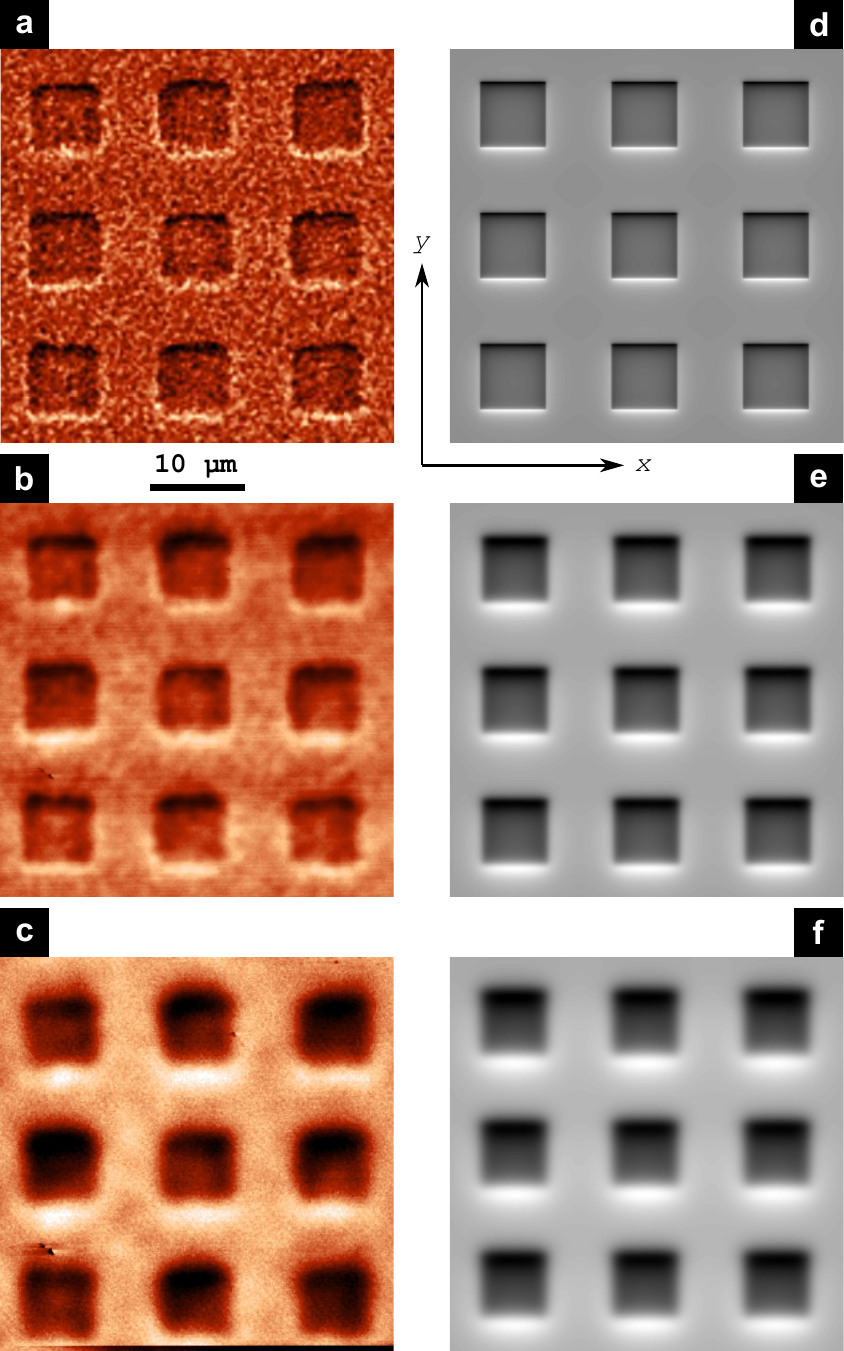}%
  \caption{\label{fig-liftHeightOnAsymmetry}(a-c)~Experimental MFM images recorded above the SmCo TMP at instrumental lift height $50$, $500$ and $\unit[1000]{\nano\metre}$, respectively~(the $\unit[100]{\nano\meter}$-thick capping layer should be added to these figures). (d-f)~Calculated maps of $H_z+H_y$, at the same heights. The contrast-to-color code is different on each image.}
  \end{center}
\end{figure}

\subfigref{fig-liftHeightOnAsymmetry}{a-c} shows MFM images of SmCo TMP performed at various heights. The qualitative variation across the images at different scan heights, with the overall contrast between the dots and matrix becoming much stronger at longer distances, may be attributed to the fact that the stray field value and thus the image contrast is strongest close to dot edges when close to the sample. This was already seen in experimental and simulated line scans in \subfigref{fig-variousHeights}{(a,b)} but at further distances, so the edge contrast was not dominant. The edge contrast around each dot is asymmetric along~$y$, as in \figref{fig-asymmetricContrast}. \subfigref{fig-liftHeightOnAsymmetry}{(d-f)} show the corresponding simulations, with a contrast  based on a linear combination of $H_{\mathrm{d},z}$ and $H_{\mathrm{d},y}$. The best agreement is found, for any fly height, for $\approx H_{\mathrm{d},z}-H_{\mathrm{d},y}$, suitable for all fly heights. This would mean $\beta=\angledeg{45}$, whereas the experimental tilt is $\beta=\angledeg{20}$. The reason for this discrepancy is not clear, either direction of moment of the probe, deviations from the model involving a spatial derivative of the stray field etc.

The qualitative agreement observed between simulations and experiments suggests that probe magnetization rigidity is fulfilled in our experiments. Note that in the case of low-coercivity probes, a four-fold symmetric phase map is achieved\bracketsubfigref{fig-asymmetricContrast}b, since the magnetisation of the tip is reversed in the stray field of the sample.

\section*{Conclusion}
\label{sec-conclusion}

In this study we examined some specific aspects of MFM when imaging permanent magnet films. Comparing commercial probes with a hard magnetic coating of nominal thickness $\unit[50]{\nano\metre}$ but different cantilever stiffness ($\approx\unit[2]{\newton\per\metre}$ and $\approx\unit[40]{\newton\per\metre}$, respectively), we have shown that because of the interplay in the first scan between the feedback and the strong probe-sample interactions, the probe with the stiffer cantilever is a better choice to make reliable topographic and quantitative magnetic images. A comparison of experimental MFM phase maps and simulations of a test sample consisting of micron-sized hard magnetic dots in a hard magnetic matrix of reverse magnetization, indicates that the MFM magnetic image represents the map of the stray field produced by the sample, as opposed to the usually reported 1st or 2nd derivative. Experimental phase maps made on out-of-plane magnetized samples using probes with a hard magnetic coating magnetized in the direction perpendicular to the cantilever, are characterized by a two-fold in-plane symmetry. We attribute this to a slight tilt of the cantilever with respect to the sample plane, so that the tip magnetization is not perpendicular to the film plane, resulting in the end in a sensitivity to the in-plane component of the samples' stray field. Our results concerning the choice of cantilever stiffness and the sensitivity to an in-plane field component are relevant for the interpretation of MFM images of hard magnet materials in both film and bulk form. Our conclusion that MFM phase maps vary as the stray field for samples with micron-scaled domains, could contribute to quantitative MFM imaging of such samples.

\section*{Annexes}
\label{sec-annexes}

\subsection*{Annexe 1: model of an AFM cantilever}
\label{sec-annexe-AFMmodel}

We recall here the usual description of a cantilever for atomic force microscopy with a harmonic oscillator along a direction with coordinate~$z$, with inertia~$m$, stiffness~$k$ and damping~$\Gamma$, subject to an external force~$F$:

\begin{equation}
  m\fracdiffn{z}{t}{2}+\Gamma\fracdiff{z}{t}+kz=F(z,t).
\end{equation}
This description is adequate for the physics of magnetic force microscopy. $F$ may contain the cantilever excitation at a given angular frequency $\omega$, as well as the force created by the sample \Fts. The latter is classically expanded to first order to remain within the frame of a harmonic oscillator. As $\linediff Fz$ is much smaller than the cantilever stiffness, this results simply in a linear shift of the resonance angular frequency from its natural value $\omegaZero = \sqrt{k/m}$, to $\omegaZero\left({1-\frac{1}{2k}\fracdiff{F}{z}}\right)$. In magnetic force microscopy, the information about the magnetic force is then mapped as the phase shift induced on the cantilever when excited at $\omega= \omegaZero$:

\begin{equation}
  \Delta\phi = - \frac{Q}{k}\fracdiff{F_\mathrm{ts}}{z}
\end{equation}
$Q=\sqrt{mk}/\Gamma$ is the quality factor of the cantilever. This linear relationship remains valid as long as the phase shift is reasonably smaller than $\pi/4$. As the force is the gradient of energy, we may rewrite the phase shift as:

\begin{equation}
  \label{eqn-phaseShiftWithEnergyCurvature}
  \Delta\phi =  \frac{Q}{k}\fracdiffn{E_\mathrm{ts}}{z}{2}
\end{equation}
The latter expression is particularly useful for the analysis of magnetic forces.

\subsection*{Annex~2: model of tip-sample magnetic interactions}
\label{sec-annexe-AFMmodel}

Formalisms for describing the tip-sample interaction in magnetic force microscopy were reported in a number of papers\cite{bib-LOH1999,bib-RUG1990,bib-PRO2005a}. We recall the basic features here, with a view to providing notations for the discussion in the body of this manuscript. We do not consider the modeling of etched bulk tips used in early MFM studies\cite{bib-HAR1989} but rather the interpretation based on sample surface charges\cite{bib-HUB1997}, which may be considered more suited for imaging surfaces of bulk materials and thick films. We also assume that both tip and sample have a rigid distribution of magnetization, \ie we do not consider mutual (susceptibility) effects. The tip-sample energy at fly height $z$ may be expressed in the most general way as:

\begin{equation}
  \label{eqn-energyTip}
  E_\mathrm{ts}(z) = -\muZero M_i\int_{-\infty}^{+\infty}H_i(z+h)s(h)\diff h
\end{equation}
where $i$ is summed over all three directions, $h$ is the coordinate along the oscillation direction, and $s(h)$ is the integrated cross-sectional area of magnetized material at height $h$ along the tip, calculated from its apex. This convoluting function was called a probe function\cite{bib-HUB1997}. In the following we assume that the tip oscillates along the vertical direction so that only spatial derivatives along $z$ need to be considered to calculate phase shifts~(see Ref.\citenum{bib-RUG1990} for the general description). $s_\mathrm{dip}(h)=v_0\delta(0)$ for the dipole model, with $\delta$ being the delta function and $v_0$ the dipole volume, while $s_\mathrm{mono}(h)=s_0 H(h)$ for the monopole model with $H$ the Heaviside function and $s_0$ the area of the equivalent cylinder of magnetic material. Following \eqnref{eqn-phaseShiftWithEnergyCurvature}, integrating \eqnref{eqn-energyTip} by parts and neglecting boundary conditions with the argument that $H$ decays towards infinity, the phase shift may be expressed in different ways based on:

\begin{eqnarray}
  \label{eqn-phaseShiftModelDipole}\fracdiffn{E_\mathrm{ts}}{z}2(z) & = & -\muZero M_i\int_{-\infty}^{+\infty}\fracdiffn{H_i}{z}2(z+h)\;s(h)\;\diff h \\
  \label{eqn-phaseShiftModelMonopole}&= & + \muZero M_i\int_{-\infty}^{+\infty}\fracdiff{H_i}{z}(z+h)\;\fracdiff{s}{z}(h)\;\diff h \\
  \label{eqn-phaseShiftModelCone}&= &-\muZero M_i\int_{-\infty}^{+\infty}H_i(z+h)\;\fracdiffn{s}{z}2(h)\;\diff h.
\end{eqnarray}
Textbook cases are those for which in one of the above equations, the probe function or one of its derivatives is a delta function. Cases usually considered in the literature are \eqnref{eqn-phaseShiftModelDipole} for a dipole and \eqnref{eqn-phaseShiftModelMonopole} for a monopole, showing in a straightforward fashion that within these models, is it the second or first $z$ derivative of the stray field at height $h$ which is being probed. This is readily seen for the latter case in \eqnref{eqn-phaseShiftModelMonopole}, for which $\linediff sz=s_0 \delta(h)$. Let us discuss the suitability of these models to describe different situations.

In the case of a very fast decay of stray field above the sample, and rapid lateral variation, only the magnetic capping at the nearly flat end of the apex is involved in the tip-sample force, and the dipole model is the best suited.

The monopole model is more appropriate for a slower vertical decay and lateral variation, typically in the range of about $\unit[100]{\nano\metre}$. In this case all the magnetic material deposited around the tip apex is involved in the interaction. Assuming that deposition is performed along the direction facing a tip with parabolic shape, it can be readily shown that in the case where the deposit thickness is smaller than the apex radius, the amount of material per unit $z$ is constant. This justifies the monopole model.

When the stray field extends way above the apex of the tip, then the predominant shape is not anymore parabolic but rather conical. To describe this case, which is the most relevant for measurements on permanent magnets, we thus consider the model of a cone capped with a uniform layer of thickness $t$ (measured along the direction perpendicular to the local surface of the cone) and half angle $\alpha$:

\begin{equation}
  s_\mathrm{cone}(h)=2\pi h t \tan\alpha.
\end{equation}
Note that within a constant, $s_\mathrm{dip}$ is the derivative of $s_\mathrm{mono}$, which itself is the derivative of $s_\mathrm{cone}$. Thus, \eqnref{eqn-phaseShiftModelCone} shows in a straightforward fashion, that in this case it is the stray field itself along the direction of magnetization of the tip that is imaged.

\subsection*{Annexe 3: expressions for calculating spatial gradients of the stray field arising from a uniformly magnetized prism}
\label{sec-annexe-AFMmodel}

Analytical formulas have long been used for calculating the stray field arising from uniformly magnetized prisms\cite{bib-RHO1954}. A concise formalism for the use of these formulas was introduced by Hubert, based on the core Green function $F_{000}=1/r$ and its space primitives $F_{ijk}$ of any order and along any direction\cite{bib-HUB1998b}. For instance, $F_{000}$ integrated once along $x$ and once along $z$, is written $F_{101}$. The scalar potential arising at any point of space from a charged plate can be expressed using these functions. The stray field arising from a uniformly-magnetized body may then be expressed using these functions, summing over its oppositely charged surfaces.

For the present case we needed to calculate first and second derivatives of the stray field arising from uniformly-magnetized prisms, here described by sets of two plates with opposite charges. For flexibility and ease of calculation we extended the notation of Hubert and calculated $F_{ijk}$ for various and indifferently positive or negative exponents. For instance, $\linediff{H_{\mathrm{d},x}}{z}$ arising from a rectangular plate in the $xy$ plane involves $F_{01-1}$: the potential involves $F_{110}$, calculating $H_{\mathrm{d},x}$ turns it into $F_{010}$, and the $z$ derivative into $F_{01-1}$. The function is then used in the following way for a plate centered on $(0,0,0)$ and of side length $a$ and~$b$:

\begin{equation}
  H_{\mathrm{d},x}(\vect r)=-\frac{\Ms}{2}\sum_{\begin{matrix}\epsilon_i=\pm1\cr\epsilon_j=\pm1\end{matrix}}\epsilon_i\epsilon_j F_{01-1}
  \left({x-\epsilon_i\frac{a}{2},y-\epsilon_j\frac{b}{2},z}\right).
\end{equation}
Below is the expression of the functions required to calculate the second spatial derivative of field arising from plates:

\begin{alignat}{7}
  &F_{1 1 -1} && = -\frac{P_z}{z} \\
  &F_{1 -1 0} && =  -\frac{xy}{r(v+w)} \\
  &F_{1 1 -2} && =  \frac{xy}{wr^2+uv}\left({r+\frac{w}{r}}\right) \\
  &F_{-1 0 0} && =  -x F_{000}^3 \\
  &F_{-1 1 -1} && =  \frac{xyz}{r^3}\;\frac{3r^2-v}{(r^2-v)^2} \\
  &F_{1 -2 0} && =  \frac{x(2v-w)r^2-xuv}{r^3(v+w)^2} \\
  \nonumber &F_{1 1 -3} && =  \frac{xyzr}{(uv+wr^2)}\\
     & && \times \left[{-\frac{2r^2}{uv+wr^2}\left({1+\frac{w}{r^2}}\right)^2 + \frac{1}{r^2}\left({3-\frac{w}{r^2}}\right) }\right].
\end{alignat}
In these expressions we use the same notations as Hubert\cite{bib-HUB1998b}:
\begin{eqnarray}
  u = x^2, v=y^2, w=z^2 \\
  P_z=z \atan(xy/zr).
\end{eqnarray}

\section*{Acknowledgements}
\label{sec-thanks}

This work was supported by the Toyota Motor Corporation and the Magnetic Materials for High-Efficient Motors (MagHEM) project.

\bibliographystyle{apsrev}

\end{document}